\documentstyle[12pt]{article}
\newcommand{\ald}{\dot \alpha}

\newcommand{\mud}{\dot \mu}

\newcommand{\ej}{\cal E}

\begin{document}
\textwidth 160mm
\textheight 240mm
\topmargin -20mm
\oddsidemargin 0pt
\evensidemargin 0pt
\newcommand{\beq}{\begin{equation}}
\newcommand{\eeq}{\end{equation}}
\begin{titlepage}
\begin{center}

\huge{\bf Gravitational SD Perturbiner}

\vspace{1.5cm}

\large{\bf A.A.Rosly and  K.G.Selivanov }

\vspace{1.0cm}

{ITEP, B.Cheremushkinskaya 25, Moscow, 117259, Russia}

\vspace{1.9cm}

{ITEP-TH-56/97, IFUM-590/FT}

\vspace{1.0cm}

\end{center}
              

\begin{abstract}
We present here an explicit self-dual classical solution of the type of
perturbiner in gravity. This solution is a generating function for
tree gravitational form-factors with all on-shell gravitons in the same 
helicity state.
\end{abstract}
\end{titlepage}

\newpage
\setcounter{equation}{0}
In this letter we continue our study of {\it perturbiners}
\cite{RS1},\cite{RS2} i.e.,
solutions of field equations (FE) which are generating functions for tree
amplitudes in the theory (more precisely - for tree form-factors,
that is ``amplitudes'' with a number of on-shell particles and one
off-shell particle in the coordinate representation). Such solution 
can be given an intrinsic definition which is formally independent
on the Feynman diagrams and which is universal in the sense that
it is applicable in any field theory. In words, the definition is as follows.
Take a linear combination of plane wave solutions of linearized
FE (with appropriate polarization factors etc.) so that
every plane wave is multiplied with a corresponding formal nilpotent
variable (a rudiment of the symbol of annihilation-creation operator).
The perturbiner is a (complex) solution of the (nonlinear) FE which is
polynomial in the given set of (nilpotent) plane waves and whose first
order part is the given solution of the linearized FE. This definition
is actually nothing but rephrasing the usual Feynman perturbation
theory for the tree form-factors{\footnote{the nilpotency
is equivalent to considering only amplitudes with no particle having
identical quantum numbers}}. Nevertheless, this definition  
 appeared to be very convenient in the cases where FE can be treated
by some other powerful methods different from the perturbation theory.
This is actually the case of the self-duality (SD) equations, both
in  gravity and in Yang-Mills (YM) theory , allowing the use of the twistor
constructions \cite{Penrose}, \cite{Ward}. The way of reducing from the
generic perturbiner to the SD one is obvious: one includes into the
plane wave solutions of the linearized FE only SD plane waves, which
is equivalent to describing only amplitudes with on-shell particles of 
a given helicity (say, of the positive one). Miraculously, the twistor
construction works extremely efficiently in the case of perturbiners,
in a sense, more efficiently than in the case of instantons. In the instanton
case it gives only explicit description of the moduli
space of solutions \cite{ADHM}, while in the perturbiner case it gives
explicit description of fields (\cite{RS2} and below).

In the case of YM theory, the idea to use the SD equations to describe
the so-called like-helicity amplitudes (see, e.g., the reviews \cite{mapa},
\cite{dixon})
was first formulated in \cite{Ba} and, independently, in \cite{Se}.
 In \cite{Ba} it was basically shown that  the SD
equations reproduce the recursion relations for tree like-helicity gluonic 
form-factors (also called ``currents''), obtained originally 
in ref.\cite{BG} from the Feynman diagrams; the  corresponding solution of 
SD equations was than obtained in terms of the known solutions 
refs.\cite{BG} of the recursion relations 
for the ``currents''. In ref.\cite{Se} an example of SD perturbiner was
obtained in the SU(2) case by a 'tHooft anzatz upon further restriction on the
on-shell particles included. A bit later an analogous solution was studied
in ref.\cite{KO}, where consideration was based
on solving  recursion relations analogous to refs. \cite{BG}.  
In \cite{RS2} the YM SD perturbiner was constructed by the twistor methods
which also allowed us to find its antiSD deformation, that is to add one
opposite-helicity gluon and thus to obtain a generating
function for the so-called maximally helicity violating Parke-Taylor 
amplitudes \cite{PT},
\cite{BG}. Also, in \cite{RS2} the SD perturbiner was constructed in
the background of an arbitrary instanton solution.

Below we present the gravitational SD perturbiner. Again, as in the YM
case, the twistor methods \cite{Penrose} give absolutely explicit
description of the fields. Thus, we present a generating function for
tree gravitational form-factors with all on-shell gravitons having the same
helicity. Derivation of the results presented as well as derivation of the 
gravitational analog of the maximally helicity violating amplitudes 
(an expression for which was conjectured in \cite{gravPT}) will be given 
elsewhere \cite{RS3}.
    
It will be convenient to use the spinor notations. To construct the perturbiner, 
one first defines the plane wave solutions
of the linearized FE. The appropriate SD solution of the linearized FE
reads ($e^{\alpha \ald}_{N}$ is a vierbein, the capital Latin letters like 
$N, M, L$ will stand for sets of particles) reads
\beq
\label{lintet}
e^{\alpha \ald}_{N}=e^{\alpha \ald}_{\emptyset}+\sum_{n{\in}N}
{\ej}_{n}\frac{q^{\alpha}_{n}\lambda^{\ald}_{n}}{(q^{n},\ae^{n})^{2}}
q_{\mu}^{n}\lambda_{\mud}^{n}dx^{\mu \mud}
\eeq
In the equation (\ref{lintet}) $e^{\alpha \ald}_{\emptyset}$ stands for the flat 
vierbein, 
$e^{\alpha \ald}_{\emptyset}=dx^{\alpha \ald}$, ${\ej}_{n}$ is the plane
wave,  ${\ej}_{n}=a_{n}e^{k^{n}_{\alpha \ald} x^{\alpha \ald}}$, $a_{n}$ is
the nilpotent symbol, $a_{n}^{2}=0$, $k^{n}_{\alpha \ald}$ is an (asymptotic)
four-momentum of the $n$-th graviton. $k^{n}$ is light-like and hence
it decomposes into a product of two spinors, 
$k^{n}_{\alpha \ald}=\ae^{n}_{\alpha}\lambda^{n}_{\ald}$.{\footnote{the reality 
of the four momentum in Minkowski space assumes that $\lambda_{\ald}=
{\bar {\ae}}_{\alpha}$}} 
As a consequence of the SD condition, the polarization tensor in 
Eq.(\ref{lintet}) contains the same dotted spinor $\lambda^{n}$ as in the 
four-momentum $k^{n}$. The other
spinor entering the polarization factor, $q^{n}_{\alpha}$, is a reference
spinor needed to define the polarization tensor, the on-shell gauge freedom
being $q^{n}_{\alpha}{\rightarrow}q^{n}_{\alpha}+{\ae}^{n}_{\alpha}$.
The factor in the
denominator is introduced for normalization, the brackets of the type of 
$(p,q)$
here and below mean contraction of two spinors with the 
$\epsilon$-tensor, $(p,q)=\epsilon^{\alpha \beta}p_{\alpha}q_{\beta}$.
$n$ numbers gravitons in the set $N$.

As it is defined in the introduction (see also \cite{RS1} and \cite{RS2}),
the perturbiner is a solution of (nonlinear) FE which is polynomial in
${\ej}_{n}, n{\in}N$  and whose linear in ${\ej}_{n}$'s part is as in 
Eq.(\ref{lintet}). We shall 
explicitly describe the corresponding vierbein $e^{\alpha \ald}_{N}$.

First we introduce auxiliary functions $u^{\ald}, \ald={\dot 1}, {\dot 2}$
which are polynomial functions of the harmonics ${\ej}_{n}, n{\in}N$
and of a pair of complex numbers $p^{\alpha}, \alpha=1,2$.
\begin{eqnarray}
\label{coord}
u^{\ald}_{N}=
\sum_{n{\in}N}\frac{(p,q_{n})^{2}{\lambda}^{\ald}_{n}}
{(p,{\ae}_{n})(q_{n},{\ae}_{n})^{2}}{\ej}_{n}
e^{(\lambda_{n},u_{N{\setminus}n}|_{p={\ae}_{n}})}\nonumber\\
=\sum_{n{\in}N}\frac{(p,q_{n})^{2}{\lambda}^{\ald}_{n}}
{(p,{\ae}_{n})(q_{n},{\ae}_{n})^{2}}{\ej}_{n}
e^{\sum_{m{\in}N{\setminus}n}
\frac{({\ae}_{n},q_{m})^{2}(\lambda_{n},\lambda_{m})}
{({\ae}_{n},{\ae}_{m})(q_{m},{\ae}_{m})^{2}}{\ej}_{m}
e^{\sum_{l{\in}N{\setminus}m,n}{\ldots}}}
\end{eqnarray}
($N{\setminus}n$ notation stands for the set $N$ without $n$th particle).
Then we can describe the vierbein $e^{\alpha \ald}_{N}$

\begin{eqnarray}
\label{pertur}
e^{\alpha \ald}_{N}=e^{\alpha \ald}_{\emptyset}
+\sum_{{\bar{M}}{\subset}N}
\frac{({\lambda}_{m_{2}},{\lambda}_{m_{1}}) \ldots
({\lambda}_{m_{M}},{\lambda}_{m_{M-1}})}
{({\ae}_{m_{2}},{\ae}_{m_{1}}) \ldots
({\ae}_{m_{M}},{\ae}_{m_{M-1}})}
e^{\sum_{m{\in}M}({\lambda}_{m},u_{N{\setminus}m}|_{p={\ae}_{m}})}\nonumber\\
{\ej}_{m_{1}} \ldots {\ej}_{m_{M}}
\frac{q^{\alpha}{\lambda}^{\ald}_{m_{1}}}{(q,{\ae}_{m_{1}})}
\frac{(q_{\mu}\lambda_{\mud}^{m_{M}}dx^{\mu \mud})}{(q,{\ae}_{m_{M}})}
\end{eqnarray}
where $\bar{M}$ is an ordered subset of $N$ and we somehow abused 
the notation because in the above formula capital $M$ stands
also for a number of particles in the set $M$.\\
Eq.(\ref{pertur}) is the sought for gravitational SD perturbiner.\\

{\it Acknowledgments}\\
Research of A.R. was partially supported by RFFI-97-02-18046. Research of K.S. was 
partially supported by Milano Section of INFN and by RFFI- 96-15-96578.
One of the authors (K.S.) thanks L.N.Lipatov for encouraging to consider perturbiner 
in the case of gravity and R.Ferrari, V.Gorini and other members of the theory group of 
Milano Section of INFN and of Milano University and H.Leutwyler and other members
of ITP at  Bern University for their hospitality while this work was completed.

\end{document}